\documentclass[
    ,final            
    ,numberedheadings 
  ]
  {aipproc}

\layoutstyle{8x11single}

\begin{document}

\title{Photometry of Some Recent Gamma-ray Bursts}

\classification{98.70.Rz, 95.85.Pw, 98.70.Qy
                }
\keywords      {gamma-ray bursts, astronomical observations, x-ray}

\author{J. K\'obori}{
  address={Konkoly Observatory}
}

\author{J. Kelemen}{
  address={Konkoly Observatory}
}
\author{P. Veres}{
  address={E\"otv\"os University, Bolyai Military University}
}
\author{B. Cenko}{
  address={CALTECH}
}
\author{D. Fox}{
  address={PSU}
}

\begin{abstract}
 We present the results of the optical, X-ray and gamma-ray analysis of some
recent GRBs. The data were obtained by the automated P60 telescope and the
Swift telescope (UVOT, XRT and BAT).  We present some example fits for the
lightcurves.  The data reduction and the investigations were made by the
Konkoly Observatory HEART group (http://www.konkoly.hu/HEART/index.html).
\end{abstract}

\maketitle


\section{Photometric data reduction}

The UVOT photometry was done using NASA's HEASOFT software package. This
package provides a complete assistance for doing photometry for measurements
done by various telescopes (e.g. Swift, CGRO, INTEGRAL).

For photometry, contrary to the suggestions of the software manual, we explored
various aperture sizes between $1$" and $10$" (instead of keeping $5$" at all times) 
to get the most usable data. Then we chose those aperture sizes, which provided 
the most accurate results.  For a given filter band we used the best aperture size, 
meaning that in some cases when obtaining photometry for a given GRB
 we used various apertures depending on the filter band.

All UVOT magnitudes are in the Standard UVOT Photometric System
\citep{uvot_phot}.  Magnitudes obtained by ground based telescopes were taken
from GCNs (in detail see at references \cite{gcn}).  When converting
magnitudes  to fluxes, we used the methods described in
\cite{bessel,p60,sdss}.  Magnitudes are not corrected for galactic extinction.
X-ray data is taken from the XRT observations available from \cite{evans}.
According to our results we suggest 19 magnitude as acceptable faintest limit
in the UVOT photometry system.

We used the \textit{mpfit} package \cite{mpfit} for fitting broken power-law
functions to the lightcurves with the $F_{\nu} \propto t^{\alpha}$ convention.

\section{Individual afterglows}
\ \ \ \underline{\textbf{GRB 080721}}:\ \
The afterglow was detected by several telescopes, but due to the lack of ground
based observations we could not fit the data at the late times. However, the
UVOT magnitudes show a lightcurve flattening about $6$ ks.  After this time the
UVOT slope in "WHITE" filter is $-0.70\pm0.18$, which is inconsistent with the
one inferred from filter R, $\alpha_R \sim -0.95$  (from $10.6$ to $26.19$
hours after the trigger), unless there was an additional steepening in the
lightcurve.  The redshift was inferred from the Lyman-$\alpha$ absorption line
and it is $z= 2.602$, while other lines (O I, Si II, C II, Si IV, C IV, Fe II
and Al II) suggest $z=2.591$.

\vspace{2mm}

\underline{\textbf{GRB 081203A}}:\ \
The $\alpha$ index in the R band disagrees with those in the GCNs
($\alpha_{R,I}=-0.66$ between $6$ and $12$ ks), but after this time the results
were similar. The redshift based on the Lyman-$\alpha$ line is about $2.1 $.

\begin{figure}[!h]
\includegraphics[width=15cm]{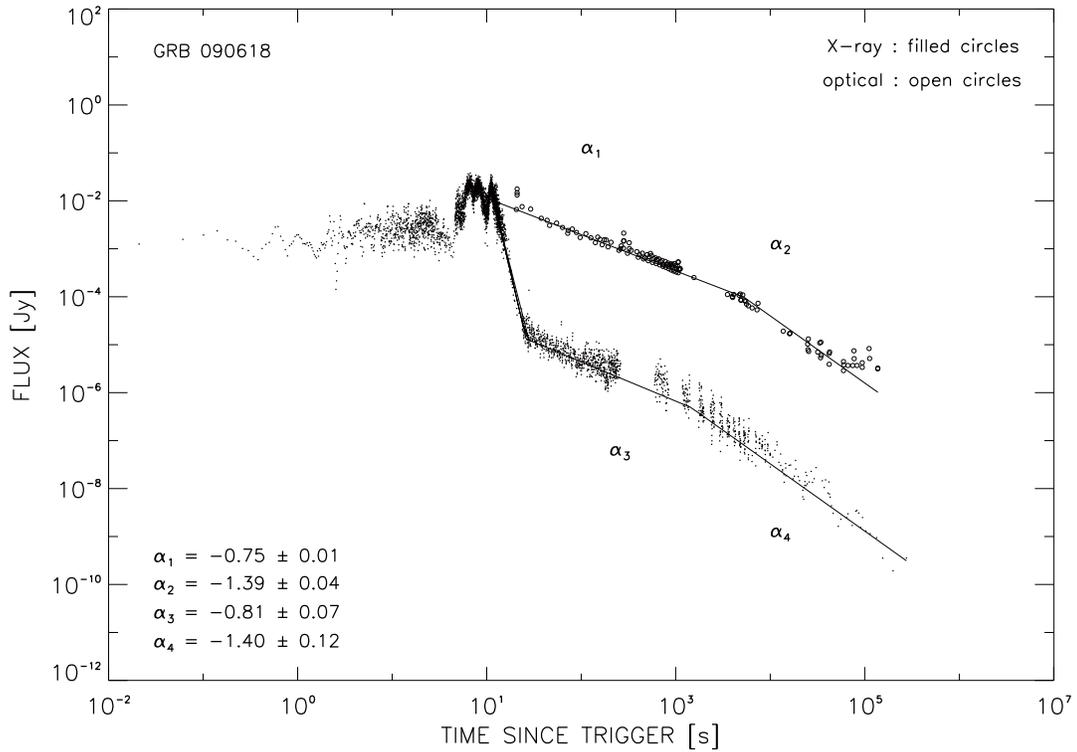}
\caption{An example of the lightcurve fitting. In the case of GRB090618 we
determined the break in the lightcurve. The open circles indicate the optical
data, which consists the following filters: V, R, I, J, H, K, r', i', z'. On
the figure, we fitted all the optical bands together to demonstrate the similar
temporal slopes in the optical and the X-ray bands.  In the R filter $2$
outlier points and  BAT measurements were entirely excluded from the fit.
 } \end{figure}

\vspace{2mm}
\underline{\textbf{GRB 090102}}:\ \
This is a well studied GRB [see GCN Circular: 8761, 8763, 8764, 8771, 8772,
8773, 8778, 8780].  Its redshift is 1.547. The observations have already begun
$43$ seconds after the BAT trigger and the TAROT magnitudes even show the
brightening interval of the lightcurve. Furthermore, in the UVOT's V and UVW1
band there is an additional rebrightening around 800 s, which can be also seen
in the X-ray band, but unfortunately, this part is lacking in ground based
observations. The lightcurve has a flattening around 1 ks.  The redshift of this
burst is $z = 1.547$.

\vspace{2mm}

\underline{\textbf{GRB 090313}}:\ \
The afterglow brightened during the first 1.3 ks, then faded until $\sim$ 10
ks.  Starting this time the plateau phase appeared, which lasted at least  until
$\sim 19$ hours after the trigger.  Then the average slope for g, r, i, z bands
changed to $\overline{\alpha} = -1.77$.  The redshift is $z = 3.375$.

\vspace{2mm}

\underline{\textbf{GRB 090618}}:\ \
This burst was very luminous with a bright afterglow. The KAIT observations
show an initial decline until 92 s then a rise at  $120 s$. After that time the
lightcurve showed two breaks: the first occurred around $600$ s, the second at
$14.6$ hours after the trigger.  However, in the UVOT observations we found a
third break (from $\overline{\alpha} = -0.64$ to $\overline{\alpha} = -1.07$),
which is in between the previous two, at $\sim 8000$ s, but ground based
telescopes did not report about such an event.  This break explains why after
the ($8000$ s) epoch the UVOT slopes did not match to the slopes reported in
GCNs (between $600$ s and $14.6$ hours $\overline{\alpha}$ = -0.76).  From RTT
images I. Khamitov et al. determined that 7.68, 8.64 and 9.58 days after the
burst the afterglow had a constant 22.3 $\pm$ 0.01 magnitude.  The redshift is
z = 0.54.  \vspace{2mm}

\underline{\textbf{GRB 090812}}:\ \
Measurements from the RAPTOR telescope system shows the lightcurve rising up to
$\sim$ 70 s, which is followed by a steady decay.  The redshift is $2.452$. 

\section{Conclusion}
\ In this work we aimed  to produce photometric data with the highest possible
accuracy.  Our sample consisted of relatively bright and well observed bursts
in order to achieve a reliable data set with our reduction method.  We excluded
those large error measurements which we had to disregard during our fitting
procedure. According to our results we suggest 19 magnitude as the faintest
limit when using the UVOT system.

\begin{theacknowledgments}
Thanks to Bob Wiegand for the help us to make HEASOFT working correctly. This
work was partially supported by OTKA grant K077795 (P.V.).
\end{theacknowledgments}





\bibliographystyle{aipproc}   


\IfFileExists{\jobname.bbl}{}
 {\typeout{}
  \typeout{******************************************}
  \typeout{** Please run "bibtex \jobname" to optain}
  \typeout{** the bibliography and then re-run LaTeX}
  \typeout{** twice to fix the references!}
  \typeout{******************************************}
  \typeout{}
 }


\begin{table}[!h]
\caption{Alpha indices. $\star$ indicates, that the slope is consistent with other measurements.
$\dag$ indicates a difference in the indices.   $\bullet$ means, that it is an average value and is taken from GCNs (in some
cases we could not fit the data, e.g. there was only the slope reported). 
Where two values are
presented, the first applies to early times, the second to the late times.}
\begin{tabular}{cccccccc}
\hline
\hline
 & GRB080721 & GRB081203A & GRB090102 & GRB090313 & GRB090618 & GRB090812 \\
\hline
&&&&&& \\
V&-1.09 $\pm$ 0.07$^{\star}$&-1.54 $\pm$ 0.03&-&-&-0.62 $\pm$ 0.04 \& & -\\
&&&&&-1.06 $\pm$ 0.10& \\
\hline
&&&&&& \\
B &-1.38 $\pm$ 0.20&-1.45 $\pm$ 0.02&-&-&-0.60 $\pm$ 0.04& -\\
&&&&&& \\
\hline
&&&&&& \\
U&-1.34 $\pm$ 0.23&-1.33 $\pm$ 0.01&-&-&-0.70 $\pm$ 0.01$^{\star}$ \& & -0.49 $\pm$ 0.10$^{\dag}$\\
&&&&&-1.04 $\pm$ 0.05& \\
\hline
&&&&&& \\
UVW1&-&-1.61 $\pm$ 0.09&-&-&-0.68 $\pm$ 0.03$^{\star}$ \& & -\\
&&&&&-1.15 $\pm$ 0.06& \\
\hline
&&&&&& \\
UVM2&-&-&-&-&-0.72 $\pm$ 0.04$^{\star}$& -\\
&&&&&& \\
\hline
&&&&&& \\
UVW2&-&-&-&-&-0.55 $\pm$ 0.06 \& & -\\
&&&&&-1.03 $\pm$ 0.04& \\
\hline
&&&&& \\
WHITE&-1.25 $\pm$ 0.01$^{\star}$ \& &-&-1.35  $\pm$ 0.14$^{\dag}$ &-&-0.73 $\pm$ 0.01 \& & -\\
&-0.70 $\pm$ 0.18&&&&-1.07 $\pm$ 0.03& \\
\hline
&&&&&& \\
R &$\sim$ -0.95$^{\bullet}$& -1.73 $\pm$ 0.01$^{\dag,\star}$ & -1.60 $\pm$ 0.02$^{\bullet}$ & -1.61 $\pm$ 0.01 & -0.69 $\pm$ 0.01$^{\star}$ & -\\
&&&$\sim$ -0.9$^{\bullet}$&&-1.27 $\pm$ 0.01$^{\star}$& \\
\hline
I &-& -1.75 $\pm$ 0.03$^{\dag,\star}$ & - & - & - &\\
\hline
J &-&- & - & -0.76 $\pm$ 0.01 & -0.83 $\pm$ 0.04 & -\\
\hline
H &-&- & - & - & -0.86 $\pm$ 0.04  & -\\
\hline
K &-&- & - & -0.55 $\pm$ 0.03  & - &-\\
\hline
g' &-& -2.27 $\pm$ 0.37 & $\sim$ -0.9$^{\bullet}$&  - & -&-1.98 $\pm$ 0.12  \\
\hline
r' &-&-& -1.00 $\pm$ 0.24 &-& -0.77 $\pm$ 0.03 & -1.24 $\pm$ 0.01\\
\hline
i' &-&-&$\sim$ -0.9$^{\bullet}$&-&-0.73 $\pm$ 0.02 & -1.21 $\pm$ 0.02$^{\star}$ \\
\hline
z' &-&-&- & - &  -0.70 $\pm$ 0.04$^{\star}$ &-\\
\hline
\hline
\end{tabular}
\end{table}

\end{document}